\documentclass[3p,times]{elsarticle}

\usepackage{ecrc}

\volume{00}

\firstpage{1}

\journalname{Nuclear Physics A}

\runauth{O. Linnyk at al.}

\jid{nupha}

\usepackage{amssymb}
\usepackage{epsfig}

\newcommand{\be}{\begin{equation}}
\newcommand{\ee}{\end{equation}}
\newcommand{\bea}{\begin{eqnarray}}
\newcommand{\eea}{\end{eqnarray}}

\begin{document}

\begin{frontmatter}



\dochead{}

\title{Dilepton production in the strongly interacting quark-gluon plasma}


\author[UF]{O.~Linnyk,\corref{cor1}}
\ead{linnyk@fias.uni-frankfurt.de} \cortext[cor1]{corresponding
author}
\author[unig]{W.~Cassing,}
\author[UF,FIAS]{E.~L.~Bratkovskaya,}
\author[unig]{J.~Manninen}

\address[UF]{
Institut f\"ur Theoretische Physik,%
Universit\"at Frankfurt am Main ,%
% Max-von-Laue Str. 1,%
 60438 Frankfurt am Main,%
 Germany%
}%
\address[unig]{
Institut f\"ur Theoretische Physik,%
  Universit\"at Giessen,%
%  Heinrich--Buff--Ring 16,%
  35392 Giessen,%
  Germany%
}
\address[FIAS]{
Frankfurt Institute for Advanced Studies,%
% Ruth-Moufang-Str. 1,%
 60438 Frankfurt am Main,%
 Germany%
}%

\begin{abstract}
Dilepton production in relativistic heavy-ion collisions is studied
within the microscopic Parton-Hadron-String Dynamics (PHSD)
transport approach, which is based on a dynamical quasiparticle
model (DQPM) matched to reproduce lattice QCD results in
thermodynamic equilibrium. A comparison to the data of the NA60
Collaboration for $In+In$ collisions at 158 A$\cdot$GeV shows
that the dilepton spectra are well described by the sum of hadronic
and partonic sources, if a collisional broadening of vector mesons
is taken into account as well as the off-shell quark-antiquark
annihilation ($q\! + \! \bar q\to l^{+}l^{-}$ and $q\! + \! \bar q
\to l^{+}l^{-} g$) in the QGP. In particular, the observed softening
of the $m_T$ spectra at intermediate masses is reproduced. The data
of the PHENIX collaboration on dilepton production in $Au+Au$ collisions
at $\sqrt{s}=200$~GeV for masses above 1~GeV are found to be
dominated by the contributions of the QGP radiation and the charm
meson decays, while the measured spectrum is underestimated in the
mass range from 0.2 to 0.6 GeV.
\end{abstract}


\end{frontmatter}

% \linenumbers

\vspace{-0.1cm}

\section{Introduction}

Dileptons are emitted over the entire space-time evolution of the
heavy-ion collision, from the initial nucleon-nucleon collisions
through the hot and dense phase and to the hadron decays after the
freeze-out. This is both a challenge and advantage of the probe. The
separation of different ``physics" in the dilepton radiation is
nontrivial due to the nonequilibrium nature of the heavy-ion
reactions and covariant transport models have to be used to
disentangle the various sources that contribute to the final
dilepton spectra seen experimentally.

To address the dilepton production in a hot and dense medium -- as
created in heavy-ion collisions -- we employ an up-to-date
relativistic transport model, i.e. the Parton Hadron String
Dynamics~\cite{CasBrat} (PHSD). The PHSD transport approach describes
the non-equilibrium evolution of relativistic heavy-ion collisions:
from the initial hard scatterings to the partonic phase in the early
hot reaction region followed by hadronization and off-shell hadron
propagation and interactions. Within PHSD, one solves generalized
transport equations on the basis of the off-shell Kadanoff-Baym
equations for Greens functions in phase-space representation (in a
first order gradient expansion beyond the quasiparticle
approximation).

The description of partons in PHSD is based on the dynamical
quasiparticle model (DQPM) matched to reproduce lattice QCD results
in thermodynamic equilibrium~\cite{Cassing06}. According to the DQPM
the constituents of the strongly interacting quark-gluon plasma
(sQGP) are massive and off-shell quasi-particles (quarks and gluons)
with broad spectral functions. In order to address the
electromagnetic radiation of the sQGP, we derived off-shell cross
sections of $q\bar q\to\gamma^*$, $q\bar q\to\gamma^*+g$ and
$qg\to\gamma^*q$ ($\bar q g\to\gamma^* \bar q$) reactions taking
into account the effective propagators for quarks and gluons from
the DQPM in~\cite{olena2010}. Dilepton production in the QGP - as created
in early stages of heavy-ion collisions - is calculated by
implementing these off-shell processes into the PHSD transport
approach.

In the hadronic sector PHSD is equivalent to the
Hadron-String-Dynamics (HSD) transport approach
\cite{CBRep98,Brat97,Ehehalt} that has been used for the description
of $pA$ and $AA$ collisions from SIS to RHIC energies and has lead
to a fair reproduction of hadron abundances, rapidity distributions
and transverse momentum spectra. In particular, HSD incorporates
off-shell dynamics for vector mesons -- according to
Refs.~\cite{Cass_off1} -- and a set of vector-meson spectral
functions~\cite{Brat08} that covers possible scenarios for their
in-medium modification.
Various models predict that hadrons change in the (hot and dense)
nuclear medium; in particular, a broadening of the spectral
functions or a mass shift of the vector mesons have been expected.
Furthermore, QCD sum rules indicated that a mass shift may lead to a
broadening and vice versa~\cite{MuellerSumRules}; therefore  both
modifications should be studied simultaneously. In the off-shell
transport description, the hadron spectral functions change
dynamically during the propagation through the medium and evolve
towards the on-shell spectral function in the vacuum.

\begin{figure*}
  \begin{minipage}[b]{0.48\textwidth}
    \includegraphics[width=\textwidth]{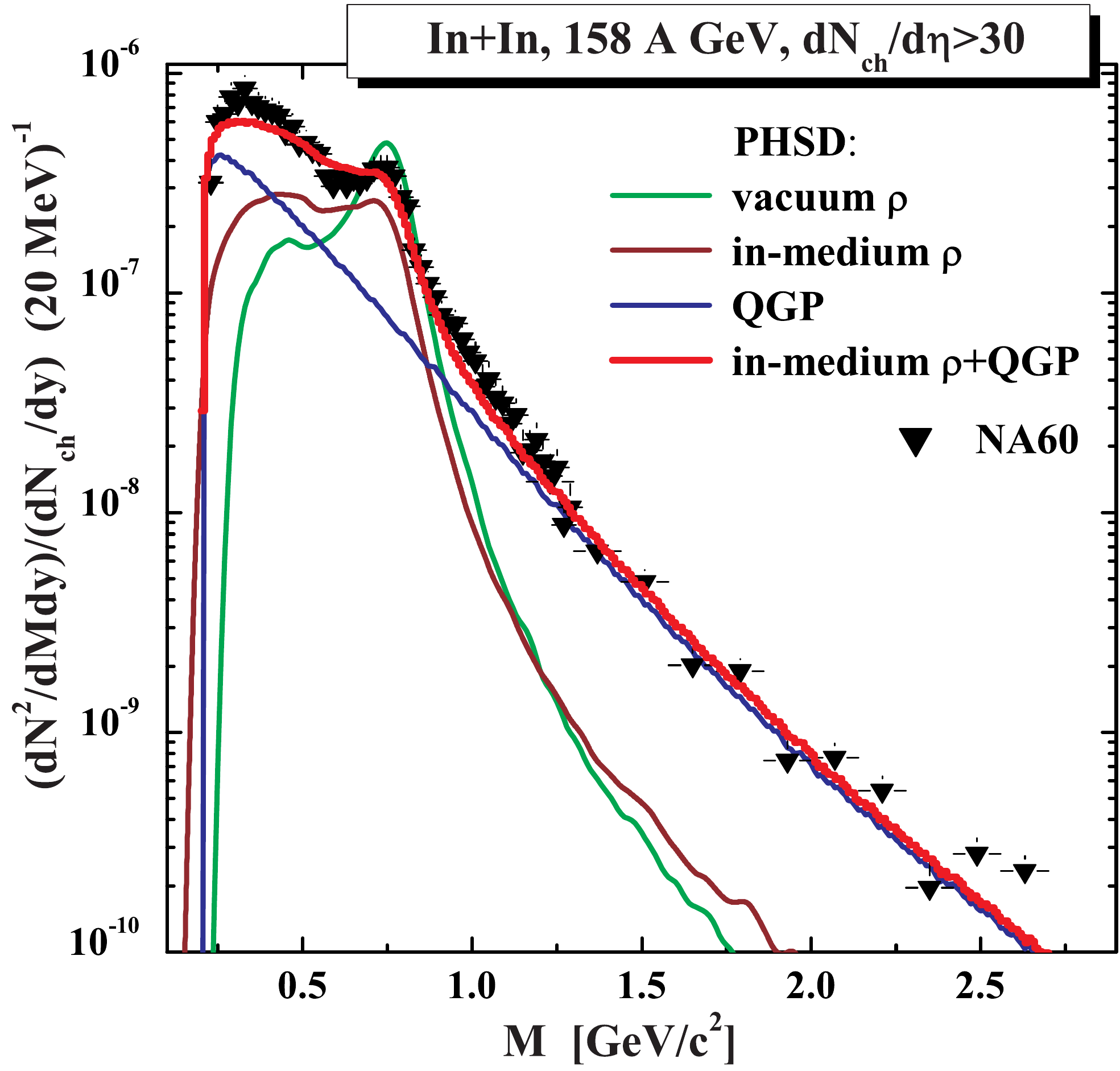}
    \label{NA60_AC}
    \caption{Acceptance corrected mass spectra of the excess dimuons from $In+In$
at 158~AGeV from PHSD compared to the data of
NA60~\cite{Arnaldi:2008er}. The green dash-dotted line shows the
dilepton yield from the vacuum $\rho$ meson. The blue dashed line is
the contribution to the dilepton yield from the in-medium $\rho$
with broadened spectral function. Red solid line presents the sum of
the in-medium $\rho$ and QGP dilepton radiation (the latter is
calculated in the on-shell approximation).}
  \end{minipage}
  \hspace{0.02\textwidth}
  \begin{minipage}[b]{0.48\textwidth}
    \includegraphics[width=\textwidth]{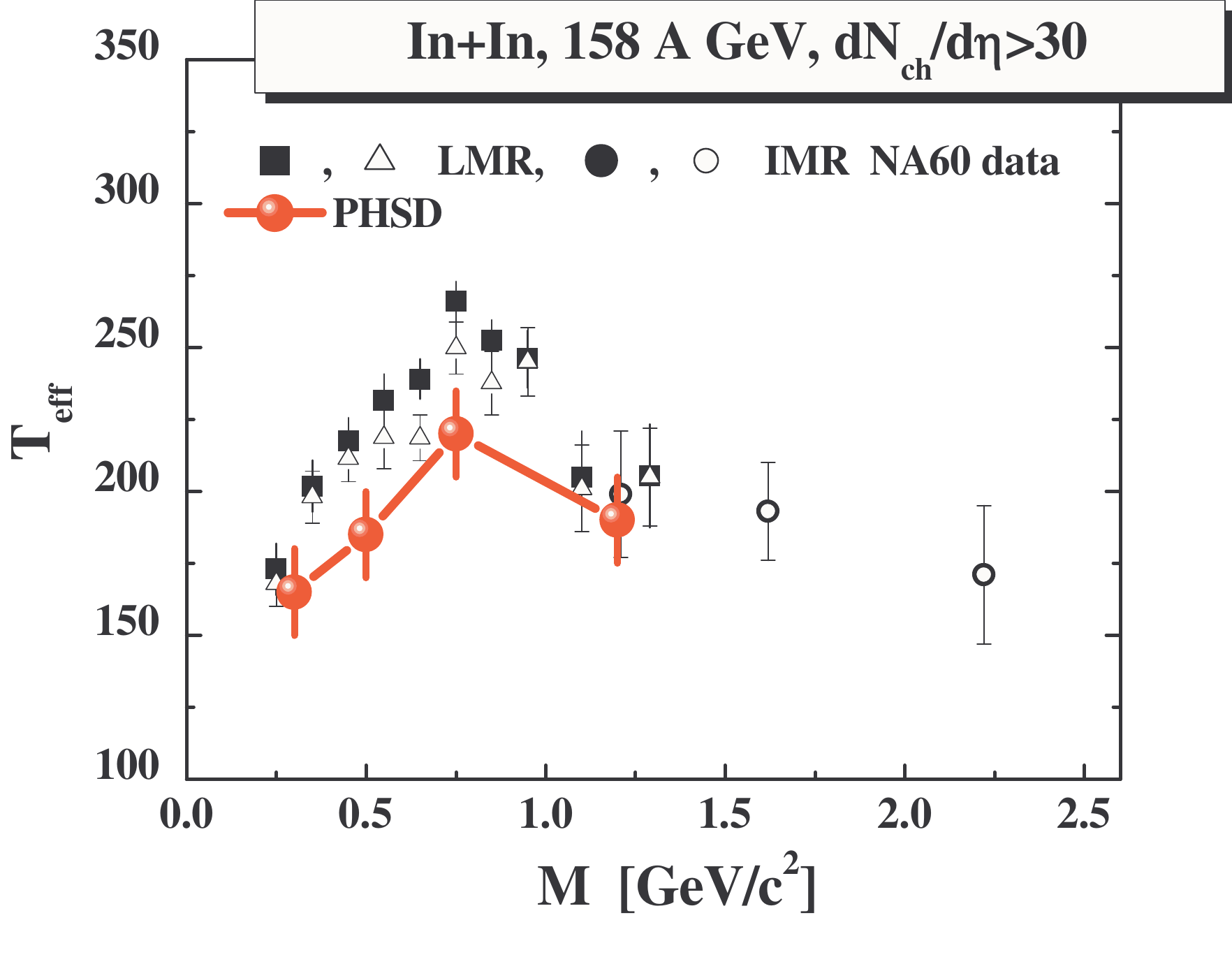}
    \label{Slopes}
    \caption{The inverse slope
parameter $T_{eff}$ of the dimuon yield from In+In at 158
A$\cdot$GeV as a function of the dimuon invariant mass in PHSD
compared to the data of the NA60
Collaboration~\protect{\cite{NA60,Arnaldi:2008er}}.}
    \vspace{1.2cm}
  \end{minipage}
\end{figure*}

\vspace{-0.1cm}

\section{Comparison to data at SPS energies}

By employing the HSD approach to the low mass dilepton production in
relativistic heavy-ion collisions, it was shown
in~\cite{here,Linnyk:2009nx,proceedings} that the NA60 Collaboration
data for the invariant mass spectra for $\mu^+\mu^-$ pairs from
In+In collisions at 158 A$\cdot$GeV favored the 'melting $\rho$'
scenario \cite{NA60,Arnaldi:2008er}. Also the data from the CERES
Collaboration \cite{CERES2} showed a preference for the 'melting
$\rho$' picture. On the other hand, the dilepton spectrum from In+In
collisions at 158 A$\cdot$GeV for $M>1$~GeV could not be accounted
for by the known hadronic sources (see Fig.2 of~\cite{here}).

The NA60 collaboration has published acceptance corrected data with
subtracted charm contribution recently~\cite{Arnaldi:2008er}. In
Fig.~\ref{NA60_AC} we present PHSD results for the dilepton spectrum
excess over the known hadronic sources as produced in $In+In$
reactions at 158~AGeV compared to the acceptance corrected data.
%
%As we find in Fig.~\ref{ExcessSpectra}, the current
The calculation in the PHSD approach confirms the earlier finding in
a hadronic model that the NA60 data favor the scenario of the
in-medium broadening of vector mesons~\cite{Linnyk:2009nx}.
Additionally, the yield at masses close to 1 GeV is reproduced by
taking into account the dilepton production from partonic channels in the QGP.
 We
find that the spectrum at invariant masses below 1~GeV is well
reproduced by the $\rho$ meson yield, if a broadening of the meson
spectral function in the medium is assumed. On the other hand, the
spectrum at $M>1$~GeV is shown to be dominated by the partonic
sources.

Moreover, accounting for partonic dilepton sources allowes to
reproduce in PHSD the effective temperature of the dileptons (slope
parameters) in the intermediate mass range~\cite{Linnyk:2009nx}, see
Fig. 2. On the other hand, most hadronic models do not
explain the softening of the $m_T$ distribution of dileptons for
$M>1$~GeV~\cite{NA60}. The softening of the transverse mass spectrum
with growing invariant mass implies that the partonic channels occur
dominantly before the collective radial flow has developed.

\vspace{-0.1cm}

\section{Comparison to data at RHIC energies}

The PHENIX Collaboration has presented dilepton data from $pp$ and
$Au+Au$ collisions at Relativistic-Heavy-Ion-Collider (RHIC)
energies of $\sqrt{s}$=200~GeV~\cite{PHENIXpp,PHENIX} which show a
large enhancement in $Au+Au$ reactions (relative to scaled $pp$
collisions) in the invariant mass regimes from 0.15 to 0.6 GeV and
from 1 to 4 GeV~\cite{Manninen:2010yf,Linnyk:2010ar}. We recall that
HSD provides a reasonable description of hadron production in
$Au+Au$ collisions at $\sqrt{s}$ = 200 GeV~\cite{Brat03}.
Whereas the total dilepton yield  is quite well described in the
region of the pion Dalitz decay as well as around the $\omega$ and
$\phi$ mass, HSD clearly underestimates the measured minimum bias
spectra in the regime from 0.2 to 0.6 GeV by approximately a factor
of 5. Between the $\phi$ and $J/\Psi$ peaks, the HSD results
underestimate the PHENIX data by approximately a factor of two.

When including the in-medium modification scenarios for the vector
mesons, we achieve a sum spectrum which is slightly enhanced
compared to the 'free' scenario.  However, the low mass dilepton
spectra from $Au+Au$ collisions at RHIC (from the PHENIX
Collaboration) are clearly underestimated in the invariant mass
range from 0.2 to 0.6 GeV in the 'collisional broadening' scenario
as well as in the 'dropping mass + collisional broadening' model. We
mention that HSD results for the low mass dileptons are very close
to the calculated spectra from van Hees and Rapp as well as Dusling
and Zahed~\cite{Dussi} (cf. the comparison in Ref.~\cite{AToia}). At
higher masses (from 1 to 4 GeV) the only hadronic sources of
correlated lepton pairs are the charmed mesons: semi-leptonic decays
of correlated D-mesons and the dilepton decays of charmonia.

\begin{figure*}
  \begin{minipage}[b]{0.48\textwidth}
    \includegraphics[width=\textwidth]{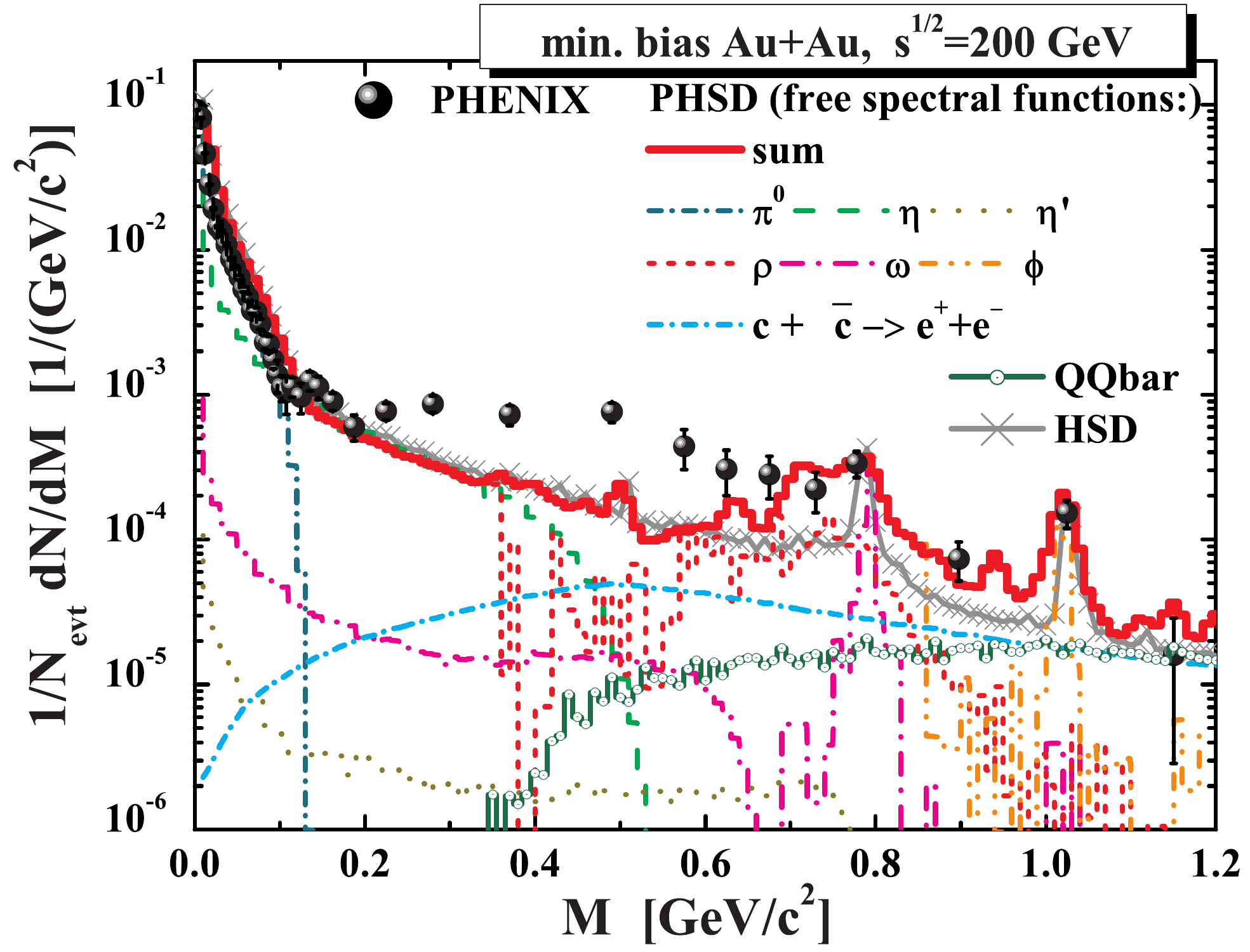}
    \caption{The PHSD results  for the mass differential dilepton spectra in case of
inclusive $Au + Au$ collisions at $\sqrt{s}$ = 200 GeV in comparison
to the data from PHENIX~\protect{\cite{PHENIX}} in the mass region
$M \! = \! 0 \! - \! 1.2$~GeV. The HSD results are shown by the grey
line with cross symbols.} \label{PHSD1}
  \end{minipage}
  \hspace{0.02\textwidth} \vspace{-0.02\textwidth}
  \begin{minipage}[b]{0.48\textwidth}
    \includegraphics[width=\textwidth]{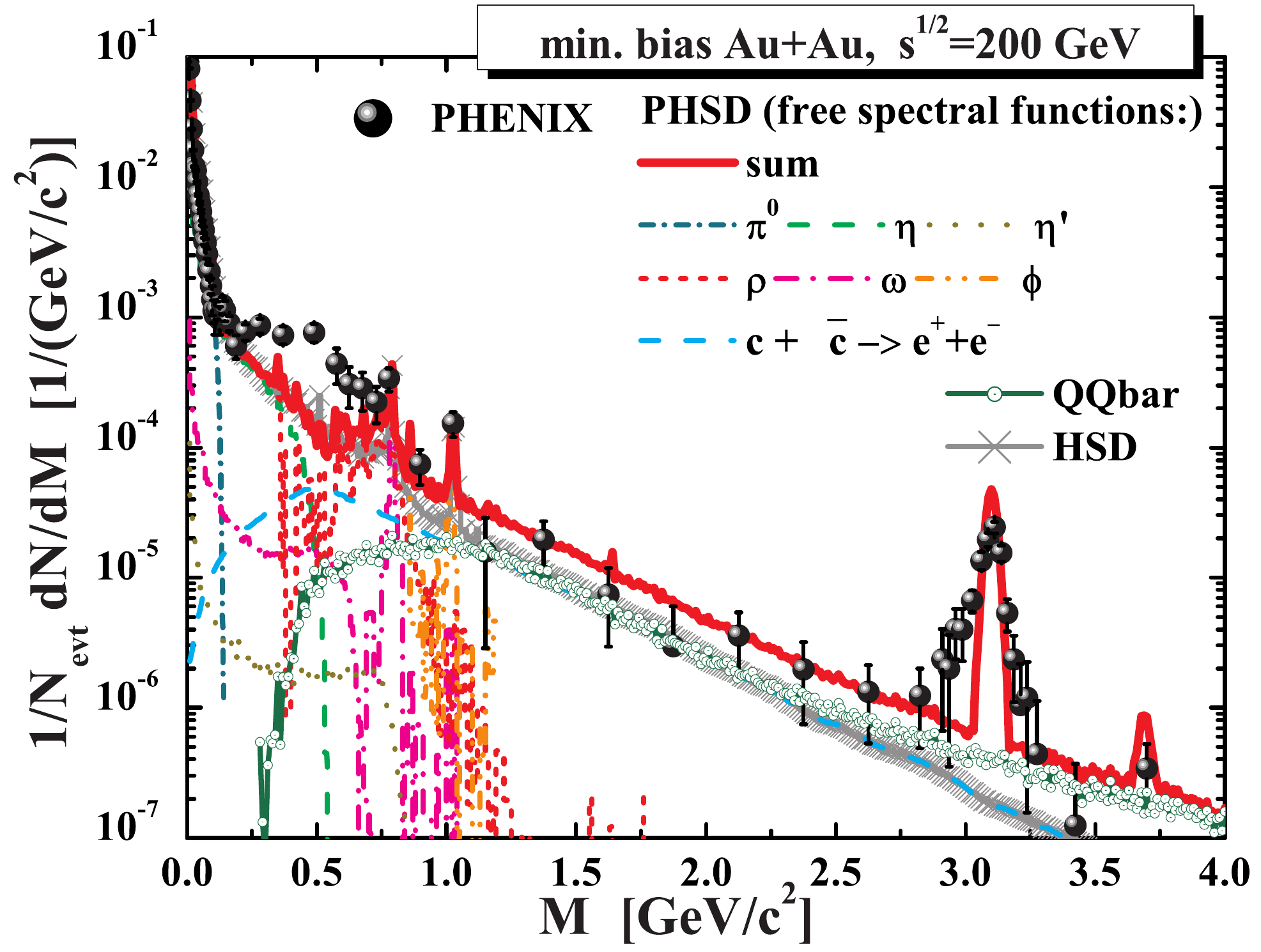}
     \caption{ The PHSD
results  for the mass differential dilepton spectra in case of
inclusive $Au + Au$ collisions at $\sqrt{s}$ = 200 GeV in comparison
to the data from PHENIX~\protect{\cite{PHENIX}} for
$M\!=\!0\!\,-\,\!4$~GeV. For comparison, the HSD results are shown
by the grey line with cross symbols.} \label{PHSD2}
  \end{minipage}
\end{figure*}

By implementing the off-shell partonic dilepton production processes
into the PHSD transport approach, we calculate the dilepton spectra
in $Au+Au$ at $\sqrt{s}$=200~GeV and compare to the PHENIX data in
Figs.~\ref{PHSD1} and \ref{PHSD2}. In Fig.~\ref{PHSD1} we present
our results for low masses ($M=0-1.2$~GeV); in this region, the
yield in PHSD is dominated by hadronic sources and essentially
coincides with the HSD result. There is a discrepancy between the
PHSD calculations and the data in the region of masses from 0.2 to
0.6 GeV. The discrepancy is not amended by accounting for the
radiation from the QGP, since the latter is subleading relative to the
radiation from hadrons integrated over the evolution of the
collision.

In Fig.~\ref{PHSD2}, the partonic radiation is visible in the mass
region $M=1-4$~GeV.  The observed yield in the mass range between
the masses of the $\phi$ and the $J/\Psi$ mesons is accounted for by
the sum of the dileptons generated by the quark-antiquark
annihilation in the sQGP and the chamed meson decays. For
$M>2.5$~GeV the partonic yield dominates over the contribution of
the (partially correlated) D-meson decays.

\vspace{-0.1cm}

\section{Summary}

The Parton-Hadron-String Dynamics (PHSD) transport approach
incorporates the relevant off-shell dynamics of the vector mesons as
well as the explicit partonic phase in the early hot and dense
reaction region. By comparing the dilepton spectrum calculated in
PHSD to the data of the NA60 and PHENIX Collaborations, we study the
relative importance of different dilepton production mechanisms and
point out the regions in phase space where partonic channels are
dominant.

A comparison of the transport calculations to the data of the NA60
Collaborations points towards a 'melting' of the $\rho$-meson at
high densities, i.e. a broadening of the vector meson's spectral
function. On the other hand, the spectrum for $M>1$~GeV is shown to
be dominated by the partonic sources.

The low mass dilepton spectra from $Au+Au$ collisions at RHIC (from
the PHENIX Collaboration) are clearly underestimated by the hadronic
channels in the invariant mass range from 0.2 to 0.6 GeV. The
discrepancy is not amended by accounting for the radiation from the
QGP, since the latter is subleading relative to the radiation from hadrons
integrated over the evolution of the collision.

In contrast, the partonic radiation is visible in the mass region
$M=1-4$~GeV. The dileptons generated by the quark-antiquark
annihilation in the sQGP constitute about half of the observed yield
in the mass range between the masses of the $\phi$ and the $J/\Psi$
mesons. For $M>2.5$~GeV the partonic yield even dominates over the
D-meson contribution. Thus, accounting for partonic radiation in
PHSD fills up the gap between the hadronic model
results~\cite{here,Manninen:2010yf} and the data for $M>1$~GeV.

%A detailed calculation of the dilepton production in heavy-ion
%collisions at RHIC energies versus collision centrality, dilepton
%mass $M$ and transverse momentum $p_T$ will be presented in near
%future. The comparison to the dilepton data will open the
%possibility to study the relative importance of different processes
%in the dilepton production and guide us towards a better
%understanding of the properties of matter created in heavy-ion
%collisions at high temperature and density.

%\section*{Acknowledgments}
%\vspace{0.2cm}

Work supported in part by the "HIC for FAIR" framework of the
"LOEWE" program and by DFG.

\vspace{-0.1cm}



%% References


%\bibliographystyle{elsarticle-num}
%\bibliography{PHSDdilept}



%\section*{References}




% \end{thebibliography}

\end{document}